\documentclass[letterpaper, 10 pt, conference]{ieeeconf}  
\IEEEoverridecommandlockouts                              
\usepackage{times}
\usepackage[pdftex]{graphicx}
\usepackage{hyperref}
\usepackage{amsmath,amsfonts,amssymb}
\usepackage{prettyref}
\usepackage{mathrsfs}
\usepackage{graphicx}
\usepackage{wrapfig}
\usepackage{xcolor}
\usepackage{subfig}
\usepackage{makecell}
\usepackage{multirow}
\usepackage{booktabs}
\usepackage{algorithm}
\usepackage{algpseudocode}
\algnewcommand{\LineComment}[1]{\State \(\triangleright\) #1}
\algdef{SE}[DOWHILE]{Do}{doWhile}{\algorithmicdo}[1]{\algorithmicwhile\ #1}
\newcommand*{\colorboxed}{}
\def\colorboxed#1#{%
  \colorboxedAux{#1}%
}
\newcommand*{\colorboxedAux}[3]{%
  \begingroup
    \colorlet{cb@saved}{.}%
    \color#1{#2}%
    \boxed{%
      \color{cb@saved}%
      #3%
    }%
  \endgroup
}

\newrefformat{Fig}{Figure~\ref{#1}}
\newrefformat{fig}{Figure~\ref{#1}}
\newrefformat{par}{Section~\ref{#1}}
\newrefformat{appen}{Appendix~\ref{#1}}
\newrefformat{sec}{Section~\ref{#1}}
\newrefformat{sub}{Section~\ref{#1}}
\newrefformat{table}{Table~\ref{#1}}
\newrefformat{alg}{Algorithm~\ref{#1}}
\newrefformat{Alg}{Algorithm~\ref{#1}}
\newrefformat{Def}{Definition~\ref{#1}}
\newrefformat{Thm}{Theorem~\ref{#1}}
\newrefformat{step}{Step~\ref{#1}}
\newrefformat{ln}{Line~\ref{#1}}
\newrefformat{eq}{Equation~\ref{#1}}
\newrefformat{pb}{Problem~\ref{#1}}
\newrefformat{it}{Item~\ref{#1}}
\newrefformat{te}{Term~\ref{#1}}
\def\Eqref Eq:#1:{\eqref{eq:#1}}
\newrefformat{Eq}{Equation~\Eqref#1:}

\newcommand{\E}[1]{\mathbf{#1}}
\newcommand{\TE}[1]{\textbf{#1}}

\newcommand{\FPP}[2]{\frac{\partial{#1}}{\partial{#2}}}
\newcommand{\FPPROW}[2]{{\partial{#1}}/{\partial{#2}}}

\newcommand{\TWO}[2]{\left(\begin{array}{cc}{#1} & {#2}\end{array}\right)}
\newcommand{\TWOC}[2]{\left(\begin{array}{c}#1 \\ #2\end{array}\right)}

\newcommand{\MTT}[4]{\left(\begin{array}{cc}#1 & #2 \\ #3 & #4\end{array}\right)}

\newcommand{\sumul}[2]{\overset{#2}{\underset{#1}{\sum}}\;}
\newcommand{\sumu}[1]{\underset{#1}{\sum}\;}
\newcommand{\argmin}[1]{\underset{#1}{\E{argmin}}\;}

\newcommand{\TWORCellC}[2]{\begin{tabular}{@{}c@{}}#1 \\ #2\end{tabular}}

\newcommand{\REFINED}[1]{{#1}}

\makeatletter
\IEEEtriggercmd{\reset@font\normalfont\fontsize{5.74pt}{5.74pt}\selectfont}
\makeatother
\IEEEtriggeratref{1}

\makeatletter

\newcommand\fs@ruled@notop{\def\@fs@cfont{\bfseries}\let\@fs@capt\floatc@ruled
  \def\@fs@pre{}%
  \def\@fs@post{\kern2pt\hrule\relax}%
  \def\@fs@mid{\kern2pt\hrule\kern2pt}%
  \let\@fs@iftopcapt\iftrue}
\renewcommand\fst@algorithm{\fs@ruled@notop}
\makeatother

\title{\vspace{-10px}\large\bf Realtime Simulation of Thin-Shell Deformable Materials 
using CNN-Based Mesh Embedding
\vspace{-15px}}
\author{Qingyang Tan$^{1}$, Zherong Pan$^{2}$, Lin Gao$^{3}$, and Dinesh Manocha$^{1}$  \\
Video Link: \url{https://youtu.be/zuXoQYJeAfc}
\vspace{-60px}
\thanks{$^1$Qingyang Tan and Dinesh Manocha are with Department of Computer Science and Electrical \& Computer Engineering, University of Maryland at College Park. \{qytan,dm@cs.umd.edu\} $^2$Zherong Pan is with Department of Computer Science, University of North Carolina at Chapel Hill. \{zherong@cs.unc.edu\} $^3$Lin Gao is with Institute of Computing Technology, Chinese Academy of Sciences. \{gaolin@ict.ac.cn\}}}

\begin{document}
\maketitle
\thispagestyle{empty}
\pagestyle{empty}

\begin{abstract}
We address the problem of accelerating thin-shell deformable object simulations by dimension reduction. We present a new algorithm to embed a high-dimensional configuration space of deformable objects in a low-dimensional feature space, where the configurations of objects and feature points have approximate one-to-one mapping. Our key technique is a graph-based convolutional neural network (CNN) defined on meshes with arbitrary topologies and a new mesh embedding approach based on physics-inspired loss term. We have applied our approach to accelerate high-resolution thin shell simulations corresponding to cloth-like materials, where the configuration space has tens of thousands of degrees of freedom. We show that our physics-inspired embedding approach leads to higher accuracy compared with prior mesh embedding methods. Finally, we show that the temporal evolution of the mesh in the feature space can also be learned using a recurrent neural network (RNN) leading to fully learnable physics simulators. After training our learned simulator runs $500-10000\times$ faster and the accuracy is high enough for robot manipulation tasks.
\end{abstract}
\section{Introduction}\label{sec:intro}
A key component in robot manipulation tasks is a dynamic model of target objects to be manipulated. Typical applications include cloth manipulation \cite{1809.08259,li2015folding}, liquid manipulation \cite{pan2016robot}, and in-hand rigid object manipulation \cite{Rajeswaran-RSS-18}. Of these objects, cloth is unique in that it is modeled as a thin-shell, i.e., a 2D deformable object embedded in a 3D workspace. To model the dynamic behaviors of thin-shell deformable objects, people typically use high-resolution meshes (e.g. with thousands of vertices) to represent the deformable objects. Many techniques have been developed to derive a dynamic model under a mesh-based representation, including the finite-element method \cite{Larson:2013:FEM:2440180}, the mass-spring system \cite{Baraff:1998:LSC:280814.280821,Choi:2002:SBR:566570.566624}, the thin-shell model \cite{Grinspun:2003:DS:846276.846284}, etc. However, the complexity of these techniques can vary from $\mathcal{O}(n^{1.5})$ to $\mathcal{O}(n^3)$ \cite{doi:10.1137/0909057}, where $n$ is the number of DOFs, which makes them very computationally cost on high-resolution meshes. For example, \cite{Narain:2012:AAR} reported an average computational time of over $1$ minute for predicting a single future state of a thin-shell mesh with around $5000$ vertices. This simulation overhead is a major cost in various cloth manipulation algorithms including \cite{1809.08259,li2015folding,lakshmanan2013constraint}.

In order to reduce the computational cost, one recent trend is to develop machine learning methods to compute low-dimensional embeddings of these meshes. Low-dimensional embeddings were original developed for applications such as image compression \cite{journals/corr/KingmaW13} and dimension reduction \cite{Zou2004}. The key idea is to find a low-dimensional feature space with approximate one-to-one mapping between a low-dimensional feature point and a high-dimensional mesh shape. So that the low-dimensional feature point can be treated as an efficient, surrogate representation of the original mesh. 

However, computing low-dimensional embeddings for general meshes poses new challenges because, unlike 2D images, meshes are represented by a set of unstructured vertices connected by edges and these vertices can undergo large distortions when cloth deforms. As a result, a central problem in representing mesh deformation data is to find an effective parameterization of the feature space that can handle arbitrary mesh topologies and large, nonlinear deformations. Several methods for low-dimensional mesh embeddings are based on PCA \cite{Alexa2000}, localized PCA \cite{Neumann:2013:SLD:2508363.2508417}, and Gaussian Process \cite{5206512}. However, these methods are based on vertex-position features and cannot handle large deformations.

\TE{Main Results:} We present a novel approach that uses physics-based constraints to improve the accuracy of low-dimensional embedding of arbitrary meshes for deformable simulation. We further present a fully learnable physics simulator of clothes in the feature space. The novel components of our algorithm include:
\REFINED{\begin{itemize}
\item A mesh embedding approach that takes into account the inertial and internal potential forces used by a physical simulator, which is achieved by introducing a physics-inspired loss function term, i.e., vertex-level physics-based loss term (PB-loss). This also preserves the material properties of the mesh.
\item A stateful, recurrent feature-space physics simulator that predicts the temporal changes of meshes in the feature space, which are modeled by introducing accurate enough for learning cloth features and training cloth manipulation controllers (see \prettyref{fig:simulateRobot}).
\end{itemize}}
To test the accuracy of our method, we construct multiple datasets by running cloth simulations using a high-resolution mesh under different material models, material parameters, and mesh topologies. We show that our embedding approach leads to better accuracy in terms of physics rule preservation than prior method \cite{Tan2018AAAI} that uses only a data term, with up to $70\%$ improvement. We have also observed up to $19\%$ and $18\%$ improvements in mesh embedding accuracy on commonly used metrics such as $\mathcal{M}_{rms}$ and $\mathcal{M}_{STED}$. Finally, we show that our feature space physics simulator can robustly predict dynamic behaviors of clothes undergoing unseen robot manipulations, while achieving $500-10000\times$ speedup over simulators running in the high-dimensional configuration space.

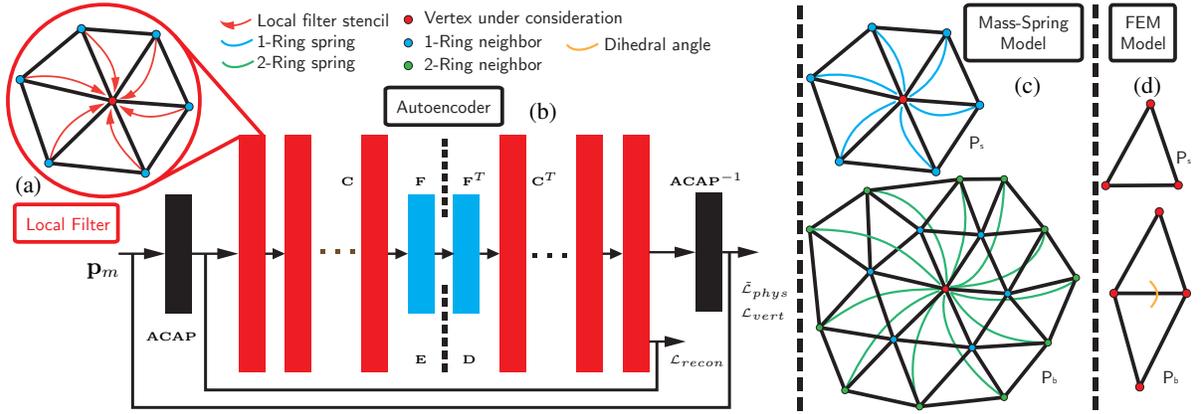
\begin{figure*}[ht]
\begin{center}
\vspace{-15px}
\scalebox{0.9}{
\includegraphics[width=0.99\textwidth]{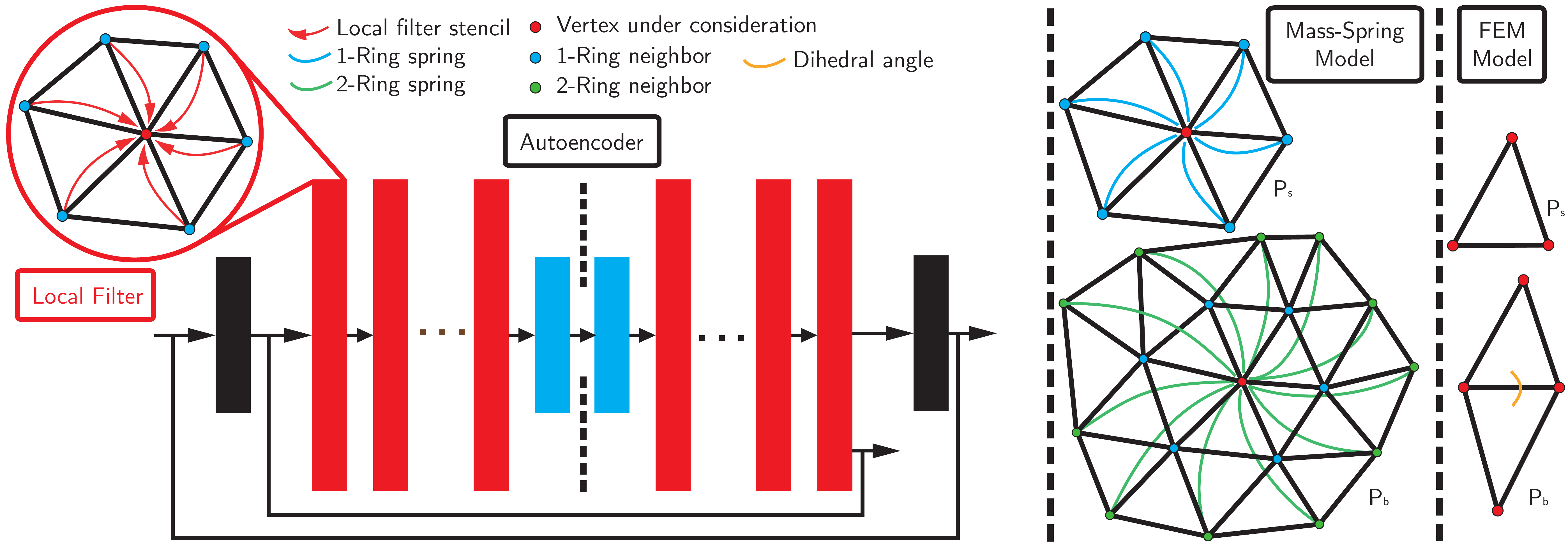}
\put(-495,92){(a)}
\put(-279,123){(b)}
\put(-75 ,134){(c)}
\put(-25 ,134){(d)}
\put(-465,58){$\E{p}_m$}
\put(-440,30){\tiny$\E{ACAP}$}
\put(-358,95){\tiny$\E{C}$}
\put(-327,95){\tiny$\E{F}$}
\put(-307,95){\tiny$\E{F}^T$}
\put(-278,95){\tiny$\E{C}^T$}
\put(-220,95){\tiny$\E{ACAP}^{-1}$}
\put(-327,20){\tiny$\E{E}$}
\put(-307,20){\tiny$\E{D}$}
\put(-220,20){\tiny$\mathcal{L}_{recon}$}
\put(-190,50){\tiny$\tilde{\mathcal{L}}_{phys}$}
\put(-190,40){\tiny$\mathcal{L}_{vert}$}}
\end{center}
\vspace{-5px}
\caption{\label{fig:method} Overview of our method: Each generated mesh $(\E{p}_m)$ is represented as vertices connected by edges. (a): We use a graph-based CNN where each convolutional layer is a local filter and the filter stencil is the 1-ring neighbor (red arrow). (b): We build an autoencoder using the filter-based convolutional layers. The decoder $\E{D}$ mirrors the encoder $\E{E}$ and both $\E{D},\E{E}$ use $L$ convolutional layers and one fully connected layer. The input of $\E{E}$ and the output of $\E{D}$ are defined in the ACAP feature space, in which we define the reconstruction loss, $\mathcal{L}_{recon}$. We recover the vertex-level features, $\E{p}_m$, using the function $\E{ACAP}^{-1}$, on which we define our PB-loss, $\mathcal{L}_{phys}$, and vertex-level regularization, $\mathcal{L}_{vert}$. The PB-loss can be formulated using two methods. (c): In the mass-spring model, the stretch resistance term is modeled as springs between each vertex and its 1-ring neighbors (blue) and the bend resistance term is modeled as springs between each vertex and its 2-ring neighbors (green). (d): FEM models the stretch resistance term as the linear elastic energy on each triangle and the bend resistance term as a quadratic penalty on the dihedral angle between each pair of neighboring triangles (yellow).}
\vspace{-10px}
\end{figure*}
The paper is organized as follows. We first review related work in \prettyref{sec:related}. We define our problem and introduce the basic method of low-dimensional mesh embedding in \prettyref{sec:VDE}. We introduce our novel PB-loss and the learnable simulation architecture in \prettyref{sec:physics}. Finally, we describe the applications in \prettyref{sec:app} and highlight the results in \prettyref{sec:results}.
\section{Related Work and Background}\label{sec:related}
We summarize related work in mesh deformations and representations, deformable object simulations, and machine learning methods for mesh deformations. 

\TE{Deformable Simulation for Robotics} are frequently encountered in service robots applications such as laundry cleaning \cite{6095109,lakshmanan2013constraint} and automatic cloth dressing \cite{Clegg:2018:LDS:3272127.3275048}. Studying these objects can also benefit the design of soft robots \cite{nakajima2017muscular,8460629}. While these soft robots are usually 3D volumetric deformable objects, we focus on 2D shell-like deformable objects or clothes. In some applications such as visual servoing \cite{Jia2018ClothMU} and tracking \cite{Brostow:2008:SRU:1478392.1478399}, deformable objects are represented using point clouds. In other applications including model-based control \cite{8460602} and reconstruction \cite{Wang:2015:DCM:2809654.2766911}, the deformable objects are represented using meshes and their dynamics are modeled by discretizing the governing equations using the finite element method (FEM). Solving the discretized governing equation is a major bottleneck in training a cloth manipulation controller, e.g., \cite{bai2016dexterous} reported up to 5 hours of CPU time spend on thin-shell simulation which is 4-5 times more costly than the control algorithm.

\TE{Deformable Object Simulations} is a key component in various model-based control algorithms such as virtual surgery \cite{alterovitz2008motion,alterovitz2009sensorless,lim2007real} and soft robot controllers \cite{8460602,duriez:hal-00823766,1809.08259}. However, physics simulators based on the finite element method \cite{Larson:2013:FEM:2440180}, the boundary-element method \cite{Brebbia:1993:IAB:187535}, or simplified models such as the mass-spring system \cite{Choi:2002:SBR:566570.566624} have a superlinear complexity. An analysis is given in \cite{doi:10.1137/0909057}, resulting in $\mathcal{O}(n^{1.5})$ complexity, where $n$ is the number of DOFs. In a high-resolution simulation, $n$ can be in the tens of thousands. As a result, learning-based methods have recently been used to accelerate physics simulations. This can be done by simulating under a low-resolution using FEM and then upsampling \cite{xie2018tempoGAN} or by learning the dynamics behaviors of clothes \cite{1802.03168} and fluids \cite{1802.10123}. However, these methods are either not based on meshes \cite{1802.10123} or not able to handle arbitrary topologies \cite{1802.03168}.

\TE{Machine Learning Methods for Mesh Deformations} has been in use for over two decades, of which most methods are essentially low-dimensional embedding techniques. Early work are based on principle component analysis (PCA) \cite{Alexa2000,Zou2004,Neumann:2013:SLD:2508363.2508417} that can only represent small, local deformations or Gaussian processes \cite{4359316,5206512} that are computationally costly to train and do not scale to large datasets. Recently, deep neural networks have been used to embed high-dimensional nonlinear functions \cite{journals/corr/KingmaW13,Radford2015UnsupervisedRL}. However, these methods rely on regular data structures such as 2D images. To handle meshes with arbitrary topologies, earlier methods \cite{7353481} represent a mesh as a 3D voxelized grid or reconstruct 3D shapes from 2D images \cite{1612.00814} using a projection layer. Recently, methods have been proposed to define CNN directly on mesh surfaces, such as CNN on parametrized texture space \cite{Maron:2017:CNN:3072959.3073616}, and CNN based on spatial filtering \cite{duvenaud2015convolutional}. The later has been used in \cite{Tan2018AAAI} to embed large-scale deformations of general meshes. Our contribution is orthogonal to these techniques and can be used to improve the embedding accuracy for any one of these methods.
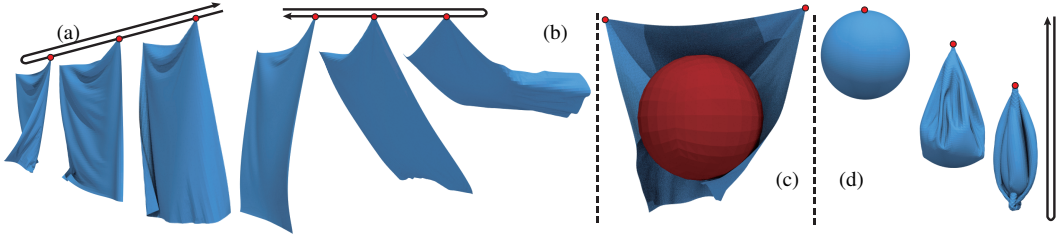
\begin{figure*}[t]
\vspace{-10px}
\begin{center}
\scalebox{0.8}{
\includegraphics[width=0.99\textwidth]{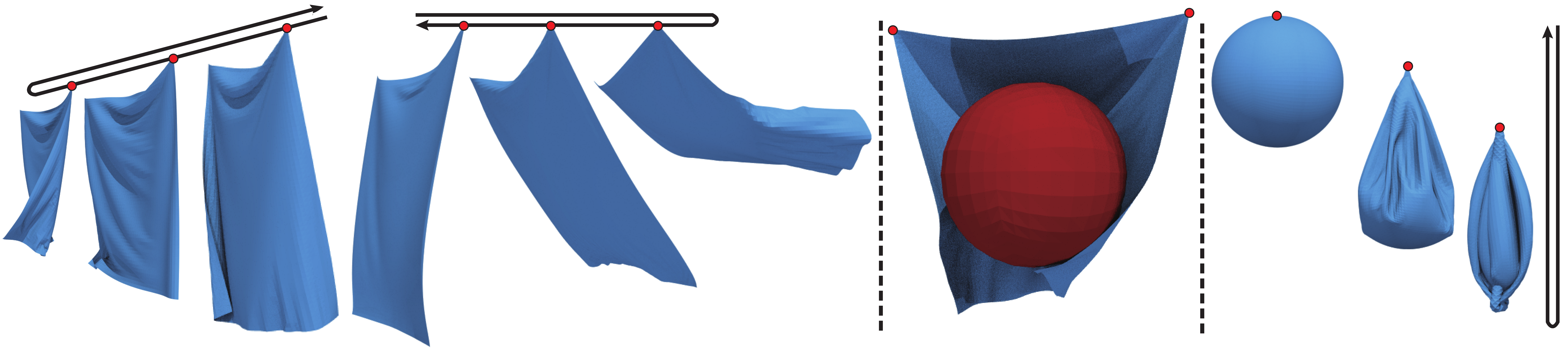}
\put(-475,94){(a)}
\put(-245,94){(b)}
\put(-135,24){(c)}
\put(-105,24){(d)}}
\end{center}
\vspace{-10px}
\caption{\label{fig:datasetVis} A visualization of our two datasets.The SHEET dataset contains $4$ simulation sequences, each with $N=2400$ frames. (a,b): We generate the dataset by grasping two corners of the cloth (red dot) and moving the grasping points back and forth along the $\pm X/Y$ axes. (c): In two sequences of the SHEET dataset, we add a spherical obstacle to interact with the cloth. (d): The BALL dataset contains $6$ simulation sequences, each with $N=500$ frames. We generate the dataset by grasping the topmost vertex of the cloth ball (red dot) and moving the grasping point back and forth along the $\pm Z$ axes.}
\vspace{-10px}
\end{figure*}

\section{Low-Dimensional Mesh Embedding}\label{sec:VDE}
In this section, we provide an overview of low-dimensional embedding of thin shell like meshes such as clothes. Our goal is to represent a set of $N$ deformed meshes, $S_m$, with each mesh represented using a set of $K$ vertices, denoted as $\E{p}_m\in\E{R}^{3K}$. We denote the $i$th vertex as $\E{p}_{m,i}\in\E{R}^3$. Here $m=1,\cdots,N$ and $i=1,\cdots,K$. These vertices are connected by edges, so we can define the 1-ring neighbor set, $\mathcal{N}^1_i$, and the 2-ring neighbor set, $\mathcal{N}^2_i$, for each $\E{p}_i$, as shown in \prettyref{fig:method} (c). Our goal is to find a map $\E{z}\to\E{p}$, where $\E{z}$ is a low-dimensional feature and $\E{p}\in\E{R}^{3K}$ such that, for each $m$, there exists a $\E{z}_m$ where $\E{z}_m$ is mapped to a mesh close to $\E{p}_m$. To define such a function, we use graph-based CNN and ACAP features \cite{Gao2017} to represent large-scale deformations.

\subsection{ACAP Feature}
For each $S_m$, an ACAP feature is computed by first finding the deformation gradient $\E{T}_{m,i}$ on each vertex:
\begin{align}
\label{eq:RECON}
\resizebox{.4\textwidth}{!}{
$\E{T}_{m,i}\triangleq\argmin{\E{T}}
\sumu{j\in \mathcal{N}^1_i}c_{ij}\|(\E{p}_{m,i}-\E{p}_{m,j}) - \E{T}(\E{p}_{1,i}-\E{p}_{1,j})\|^2,$}
\end{align}
where $c_{ij}$ are cotangent weights \cite{Desbrun:1999:IFI:311535.311576}. Here, we use $S_1$ as a reference shape. Next, we perform polar decomposition to compute $\E{T}_{m,i}=\E{R}_{m,i}\E{S}_{m,i}$ where $\E{R}_{m,i}$ is orthogonal and $\E{S}_{m,i}$ is symmetric. Finally, $\E{R}_{m,i}$ is transformed into log-space in an as-consistent-as-possible manner using mixed-integer programming. The final ACAP feature is defined as $\E{ACAP}_{m,i}\triangleq\{\E{log}(\E{R}_{m,i}),\E{S}_{m,i}\}\in\E{R}^9$ due to the symmetry of $\E{S}_{m,i}$. We denote the ACAP feature transform as: $\E{ACAP}(\E{p}_m)\in\E{R}^{9K}$. It is suggested, e.g., in \cite{5557868}, that mapping $\E{z}_m$ to the ACAP feature space leads to better effectiveness in representing large-scale deformations. Therefore, we define our mapping function to be $\E{D}(\E{z}): \E{z}\to\E{ACAP}(\E{p})$ and then recover $\E{p}$ via the inverse feature transform: $\E{ACAP}^{-1}$.

\subsection{Graph-Based CNN for Feature Embedding}
The key idea in handling arbitrary mesh topologies is to define $\E{D}$ as a graph-based CNN using local filters \cite{duvenaud2015convolutional}:
\begin{align*}
\E{D}\triangleq\E{C}_L^T\cdots\E{C}_2^T\circ\E{C}_1^T\circ\E{F}^T,
\end{align*}
where $L$ is the number of convolutional layers and $\E{C}^T$ is the transpose of a graph-based convolutional operator. Finally, $\E{F}$ is a fully connected layer. Each layer is appended by a leaky ReLU activation layer. A graph-based convolutional layer is a linear operator defined as:
\begin{align*}
\resizebox{.45\textwidth}{!}{
$\E{C}(\E{ACAP}(\E{p}_m))_{m,i}\triangleq
\E{W}\times\E{ACAP}_{m,i}+
\E{W}_\mathcal{N}\sumu{j\in \mathcal{N}^1_{i}}\E{ACAP}_{m,j}/|\mathcal{N}^1_{i}|+\E{b},$}
\end{align*}
where $\E{W},\E{W}_\mathcal{N},\E{b}$ are optimizable weights and biases, respectively. All the weights in the CNN are trained in a self-supervised manner using an autoencoder and the reconstruction loss:
\begin{align*}
\resizebox{.42\textwidth}{!}{$
\mathcal{L}_{recon}=\sumul{m=1}{N}\|\E{D}\circ\E{E}\circ\E{ACAP}(\E{p}_m)-\E{ACAP}(\E{p}_m)\|^2/N,$}
\end{align*}
\REFINED{where $\E{E}$ is a mirrored encoder of $\E{D}$ with a weight-tied architecture defined as:
\begin{align*}
\E{E}\triangleq\E{F}\circ\E{C}_1\cdots\E{C}_{L-1}\circ\E{C}_L,
\end{align*}
which means that each layer in $\E{E}$ is a transpose of the corresponding layer in $\E{D}$ with shared weights.} The construction of this CNN is illustrated in \prettyref{fig:method} (a). In the next section, we extend this framework to make it aware of physics rules.
\section{Physics-Based Loss Term}\label{sec:physics}
We present a novel physics-inspired loss term that improves the accuracy of low-dimensional mesh embedding. Our goal is to combine physics-based constraints with graph-based CNNs, where our physics-based constraints take a general form and can be used with any material models such as FEM \cite{Narain:2012:AAR} and mass-spring system \cite{Choi:2002:SBR:566570.566624}. We assumes that $S_m$ is generated using a physics simulator that solves a continuous-time PDE of the form:
\begin{align}
\label{eq:PDE}
\E{M}\FPP{\E{p}}{t}=-\E{force}(\E{p},\E{q}),
\end{align}
where $\E{M}$ is the mass matrix and $t$ is the time. This form of governing equation is the basis for state-of-the-art thin shell simulators including \cite{Choi:2002:SBR:566570.566624,Narain:2012:AAR}. $\E{force}(\E{p},\E{q})$ models internal and external forces affecting the current mesh $\E{p}$. The force is also a function of the current control parameters $\E{q}$, which are the positions of the grasping points on the mesh (red dots of \prettyref{fig:datasetVis}). This continuous time PDE \prettyref{eq:PDE} can be discretized into $N$ timesteps such that $S_m$ is the position of $S$ at time instance $i\Delta t$, where $\Delta t$ is the timestep size. A discrete physics simulator can determine all $\E{p}_m$ given the initial condition $\E{p}_1,\E{p}_2$ and the sequence of control parameters $\E{q}_1,\E{q}_2,\cdots,\E{q}_N$ by the recurrent function:
\begin{align}
\label{eq:SIM}
\E{p}_m\triangleq f(\E{p}_{m-2},\E{p}_{m-1},\E{q}_{m}),
\end{align}
where $f$ is a discretization of \prettyref{eq:PDE}. To define this discretization, we use a derivation of \cite{Liu:2013:FSM} that reformulates $f$ as the following optimization:
\begin{align}
\label{eq:PLOSS}
\E{p}_m&\triangleq\argmin{\E{p}}\mathcal{L}_{phys}(\E{p}_{m-2},\E{p}_{m-1},\E{p},\E{q}_{m})
\end{align}
\begin{align*}
\mathcal{L}_{phys}&\triangleq\|\E{p}-2\E{p}_{m-1}+\E{p}_{m-2}\|_\E{M}^2/(2\Delta t^2)+\mathcal{P}(\E{p},\E{q}_m)\nonumber.
\end{align*}
Note that \prettyref{eq:PLOSS} is just one possible implementation of \prettyref{eq:SIM}. Here the first term models the kinematic energy, which requires each vertex to move in its own velocity as much as possible if no external forces are exerted. The second term models forces caused by various potential energies at configuration $\E{p}$. In this work, we consider three kinds of potential energy: 
\begin{itemize}
    \setlength{\parskip}{-1.0pt}
    \item Gravitational energy $\mathcal{P}_g(\E{p})\triangleq-\sumul{i=1}{K}\E{g}^T\E{M}\E{p}$, where $\E{g}$ is the gravitational acceleration vector.
    \item Stretch resistance energy, $\mathcal{P}_s$, models the potential force induced by stretching the material.
    \item Bending resistance energy, $\mathcal{P}_b$, models the potential force induced by bending the material.
\end{itemize}
There are many ways to discretize $\mathcal{P}_s,\mathcal{P}_b$, such as the finite element method used in \cite{Narain:2012:AAR} or the mass-spring model used in \cite{Liu:2013:FSM,Choi:2002:SBR:566570.566624}. Both formulations are evaluated in this work. 
\begin{itemize}
    \setlength{\parskip}{-1.0pt}
    \item \cite{Choi:2002:SBR:566570.566624} models the stretch resistance term, $\mathcal{P}_s$, as a set of Hooke's springs between each vertex and vertices in its 1-ring neighbors. In addition, the bend resistance term, $\mathcal{P}_b$, is defined as another set of Hooke's springs between each vertex and vertices in its 2-ring neighbors. (\prettyref{fig:method} (c))
    \item \cite{Narain:2012:AAR} models the stretch resistance term, $\mathcal{P}_s$, as a linear elastic energy resisting the in-plane deformations of each mesh triangle. In addition, the bend resistance term, $\mathcal{P}_b$, is defined as a quadratic penalty term resisting the change of the dihedral angle between any pair of two neighboring triangles. (\prettyref{fig:method} (d))
\end{itemize}

Our approach uses \prettyref{eq:PLOSS} as an additional loss function for training $\E{D},\E{E}$. Since \prettyref{eq:PLOSS} is used for data generation, using it for mesh deformation embedding should improve the accuracy of the embedded shapes. However, there are two inherent difficulties in using $\mathcal{P}$ as an loss function. First, $\mathcal{P}$ is defined on the vertex level as a function of $\E{p}_m$, not on the feature level as a function of $\E{ACAP}(\E{p}_m)$. To address this issue, we use the inverse function $\E{ACAP}^{-1}$ to reconstruct $\E{p}_m$ from $\E{ACAP}(\E{p}_m)$. The implementation of $\E{ACAP}^{-1}$ is introduced in \prettyref{sec:IssueA}. By combining $\E{ACAP}^{-1}$ with $\mathcal{L}_{phys}$, we can train the mesh deformation embedding network using the following loss:
\begin{align}
&\tilde{\mathcal{L}}_{phys}\triangleq\mathcal{L}_{phys}(\E{ACAP}^{-1}\circ\E{D}(\E{z}_{m, m-1, m-2}), \E{q}_m)\nonumber   \\
&\mathcal{L}_{ephys}=\tilde{\mathcal{L}}_{phys}(\E{E}\circ\E{ACAP}(\E{p}_{m,m-1,m-2}),\E{q}_m).
\label{eq:embedding_phys}
\end{align}
Our second difficulty is that the embedding network is stateless and does not account for temporal information. In other words, function $\E{E}$ only takes $\E{p}_m$ as input, while \prettyref{eq:PLOSS} requires $\E{p}_m,\E{p}_{m-1},\E{p}_{m-2}$. To address this issue, we use a small, fully connected, recurrent network to represent the physics simulation procedure in the feature space. The training of this stateful network is introduced in \prettyref{sec:IssueB}. Finally, in addition to the PB-loss, we also add an autoencoder reconstruction loss on the vertex level as a regularization:
\begin{align*}
\resizebox{.45\textwidth}{!}{$
\mathcal{L}_{vert}=\sumul{m=1}{N}\|\E{ACAP}^{-1}\circ\E{D}\circ\E{E}\circ\E{ACAP}(\E{p}_m)-\E{p}_m\|^2/N.$}
\end{align*}
\REFINED{This vertex level loss can be removed from our loss function without significantly affect the quality of results. However, $\mathcal{L}_{vert}$ provides complementary information to $\mathcal{L}_{recon}$. Since ACAP is feature in the gradient domain, using only $\mathcal{L}_{recon}$ will reconstruct accurate local geometric features, but can lead to large error in vertices' positions. Therefore, we combine $\mathcal{L}_{vert}$ and $\mathcal{L}_{recon}$ to reduce the gradient domain errors and absolute vertex position errors.}

\subsection{The Inverse of the ACAP Feature Extractor\label{sec:IssueA}}
The inverse of the $\E{ACAP}$ function (black block in \prettyref{fig:method}) involves three steps. Fortunately, each step can be easily implemented in a modern neural network toolbox such as TensorFlow \cite{45381}. The first step computes $\E{R}_{m,i}$ from $\E{Log}(\E{R}_{m,i})$ using the Rodrigues' rotation formula, which involves only basic mathematical functions such as dot-product, cross-product, and the cosine function. The second step computes $\E{T}_{m,i}$ from $\E{R}_{m,i},\E{S}_{m,i}$, which is a matrix-matrix product. The final step computes $\E{p}_{m,i}$ from $\E{T}_{m,i}$. According to \prettyref{eq:RECON}, this amounts to pre-multiplying the inverse of a fixed sparse matrix, $\E{L}$, representing the Poisson reconstruction. However, this $\E{L}$ is rank-3 deficient because it is invariant to rigid translation. Therefore, we choose to define a pseudo-inverse by fixing the position of the grasping points $\E{q}$:
\begin{align}
\label{eq:invACAP}
\E{L}^\dagger\E{p}\triangleq\TWO{\E{I}}{\E{0}}\MTT{\E{L}}{\E{A}^T}{\E{A}}{\E{0}}^{-1}\TWOC{\E{p}}{\E{q}},
\end{align}
which can be pre-factorized. Here $\E{A}^{3\times3K}$ is a matrix selecting the grasping points.

\subsection{Stateful Recurrent Neural Network\label{sec:IssueB}}
A physics simulation procedure is Markovian, i.e. current configuration $\TWO{\E{p}_{m-1}}{\E{p}_{m}}$ only depends on previous configuration $\TWO{\E{p}_{m-2}}{\E{p}_{m-1}}$ of the mesh. As a result, $\mathcal{L}_{phys}$ is a function of both $\E{p}_{m-2}$, $\E{p}_{m-1}$, and $\E{p}_{m}$, which measures the violation of physical rules. However, our embedding network is stateless and only models $\E{p}_m$. In order to learn the entire dynamic behavior, we augment the embedding network with a stateful, recurrent network represented as a multilayer perceptron (MLP). This MLP represents a physically correct simulation trajectory in the feature space and is also Markovian, denoted as:
\begin{align}
\label{eq:MLP}
\E{MLP}(\E{z}_{m-2},\E{z}_{m-1},\E{q}_m)=\E{z}_m.
\end{align}
Here the additional control parameters $\E{q}$ are given to $\E{MLP}$ as additional information. 
We can build a simple reconstruction loss below to optimize $\E{MLP}$:
\begin{align*}
\resizebox{.45\textwidth}{!}{$
\mathcal{L}_{sim} = \sumul{m=3}{N}\|\E{MLP}(\E{z}_{m-2},\E{z}_{m-1},\E{q}_m)-\E{z}_m\|^2/(N-2).$}
\end{align*}
In addition, we can also add PB-loss to train this MLP, for which we define ${\mathcal{L}}_{mphys}$ on a sequence of $N$ meshes by unrolling the recurrent network:
\begin{align}
\label{eq:LPHYS}
&\mathcal{L}_{mphys} = \frac{1}{N-2}* \\
(&\tilde{\mathcal{L}}_{phys}(\E{z}_1,\E{z}_2,\E{MLP}(\E{z}_1,\E{z}_2,\E{q}_3),\E{q}_3)+\nonumber   \\
&\tilde{\mathcal{L}}_{phys}(\E{z}_2,\E{z}_3,\E{MLP}(\E{z}_2,\E{z}_3,\E{q}_4),\E{q}_4)+\cdots+\nonumber   \\
&\tilde{\mathcal{L}}_{phys}(\E{z}_{N-2},\E{z}_{N-1},\E{MLP}(\E{z}_{N-2},\E{z}_{N-1},\E{q}_N),\E{q}_N))\nonumber.
\end{align}
However, we argue that \prettyref{eq:LPHYS} will lead to a physically incorrect result and cannot be directly used for training. To see this, we note that \prettyref{eq:PLOSS} is the variational form of \prettyref{eq:PDE}. So that $\E{p}_m$ is physically correct when $\mathcal{L}_{phys}$ is at its local minima, i.e. the following partial derivative vanishes:
\begin{align}
\label{eq:VIO_MEASURE}
\FPP{\mathcal{L}_{phys}(\E{p}_{m-2},\E{p}_{m-1},\E{p}_m,\E{q}_m)}{\E{p}_m}=0\quad\forall m.
\end{align} 
However, if we sum up $\mathcal{L}_{phys}$ over a sequence of $N$ meshes and require the summed-up loss to be at a local minimum, as is done in \prettyref{eq:LPHYS}, then we are essentially requiring the following derivatives to vanish:
\begin{align}
\label{eq:VIO_MEASURE_SUM}
&\FPP{\mathcal{L}_{phys}(\E{p}_{m-2},\E{p}_{m-1},\E{p}_m,\E{q}_m)}{\E{p}_m}+\nonumber    \\
&\FPP{\mathcal{L}_{phys}(\E{p}_{m-1},\E{p}_{m},\E{p}_{m+1},\E{q}_m)}{\E{p}_m}+\nonumber    \\
&\FPP{\mathcal{L}_{phys}(\E{p}_{m},\E{p}_{m+1},\E{p}_{m+2},\E{q}_m)}{\E{p}_m}=0\quad\forall m.
\end{align}

The difference between \prettyref{eq:VIO_MEASURE} and \prettyref{eq:VIO_MEASURE_SUM} is the reason that \prettyref{eq:LPHYS} gives an incorrect result. To resolve the problem, we slightly modify the back propagation procedure of our training process by setting the partial derivatives of $\mathcal{L}_{phys}$ with respect to its first two parameters to zero:
\begin{align*}
&\FPP{\mathcal{L}_{phys}(\E{p}_{m-2},\E{p}_{m-1},\E{p}_m,\E{q}_m)}{\left[\E{p}_{m-1},\E{p}_{m-2}\right]}=0\quad\forall m,
\end{align*}
which, combined with \prettyref{eq:VIO_MEASURE_SUM}, leads to \prettyref{eq:VIO_MEASURE}. (We add similar gradient constraints when optimizing over \prettyref{eq:embedding_phys}.) This procedure is equivalent to an alternating optimization procedure, where we first compute a sequence of feature space coordinates, $\E{z}_m$, using the recurrent network (\prettyref{eq:MLP}) and then fix the first two parameters $\E{z}_{m-2},\E{z}_{m-1}$ and optimize $\mathcal{L}_{mphys}$ with respect to its third parameter $\E{z}_m$.

\section{Applications}\label{sec:app}
The two novel components in our method, the $\E{ACAP}^{-1}$ operator and the stateful PB-loss, enable a row of new applications, including realtime cloth inverse kinematics and feature space physics simulations. 

\subsection{Cloth Inverse Kinematics}
\begin{figure}[th]
\vspace{-10px}
\begin{center}
\scalebox{0.8}{
\includegraphics[width=0.45\textwidth,height=0.2\textwidth]{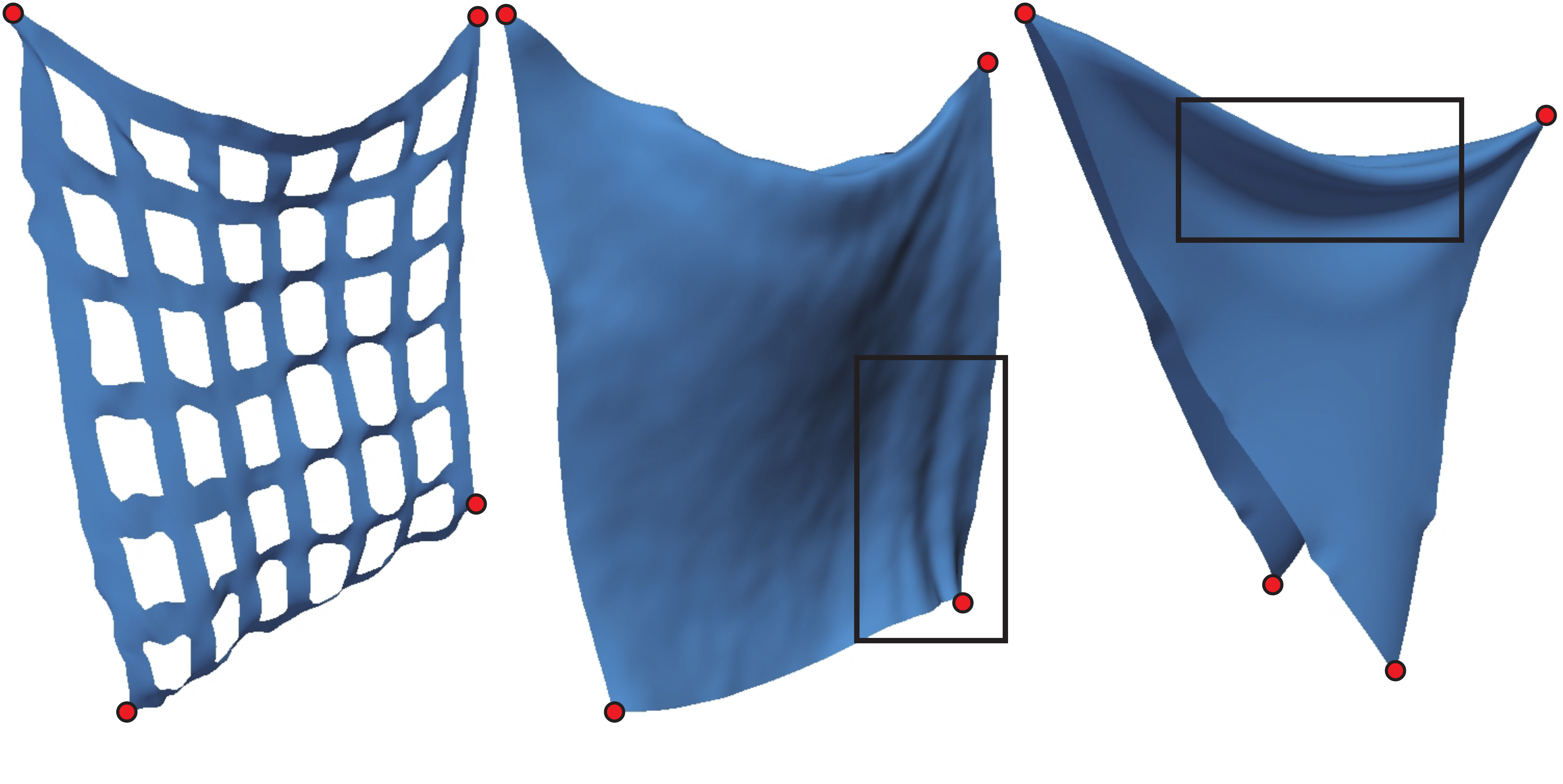}}
\end{center}
\vspace{-15px}
\caption{\label{fig:control} Three examples of cloth inverse kinematics with fixed vertices marked in red. Note that our method can synthesize detailed wrinkles and cloth-like deformations in unconstrained parts of the meshes (black box).}
\vspace{-10px}
\end{figure}

Our first application allows a robot to grasp several points of a piece of cloth and then infer the full kinematic configuration of the cloth. Such inverse kinematics can be achieved by minimizing a high-dimensional nonlinear potential energy, such as ARAP energy \cite{Sorkine:2007:ASM:1281991.1282006}, which is computationally costly. Using the inverse of the ACAP feature extractor, our method allows vertex-level constraints. Therefore, we can perform solve for the cloth configuration by a fast, low-dimensional minimization in the feature space as:
\begin{align*}
&\E{z}^*=\argmin{\E{z}}\mathcal{L}\circ\E{ACAP}^{-1}\circ\E{D}(\E{z}),
\end{align*} 
where we treat all the grasped vertices as control parameters $\E{q}$ used in \prettyref{eq:invACAP}. This application is stateless and the user controls a single feature of a mesh, $\E{z}$, so that we drop the kinetic term in $\mathcal{L}_{phys}$ and only retain the potential term $\mathcal{P}_s+\mathcal{P}_b$. Some inverse kinematic examples generated using this formulation are shown in \prettyref{fig:control}. Note that detailed wrinkles and cloth-like deformations are synthesized in unconstrained parts of the meshes.

\subsection{Feature Space Physics Simulation}
\begin{figure}[th]
\vspace{-8px}
\begin{center}
\scalebox{0.8}{
\begin{tabular}{c}
\includegraphics[width=0.45\textwidth]{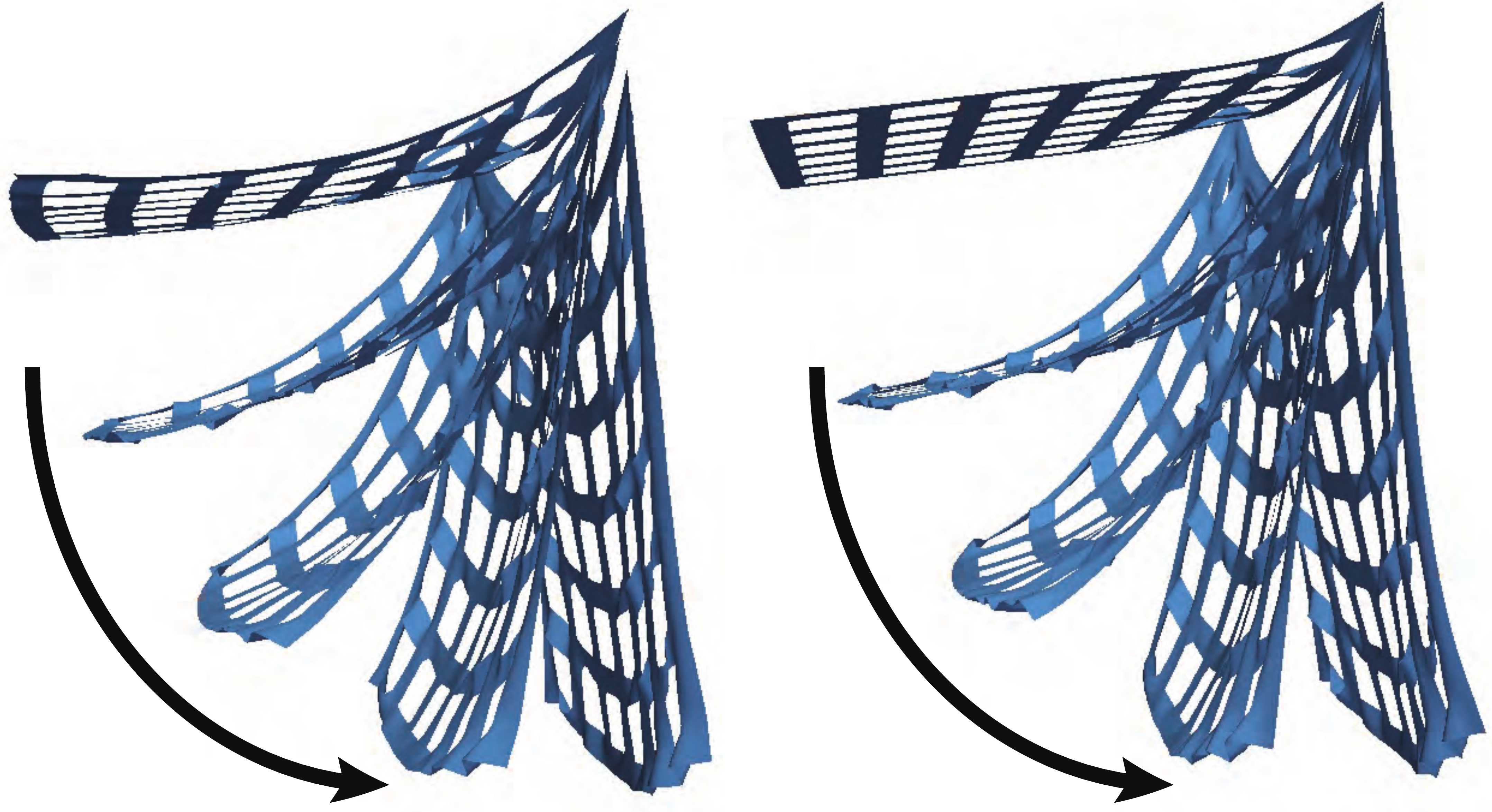}    \\
\includegraphics[width=0.45\textwidth]{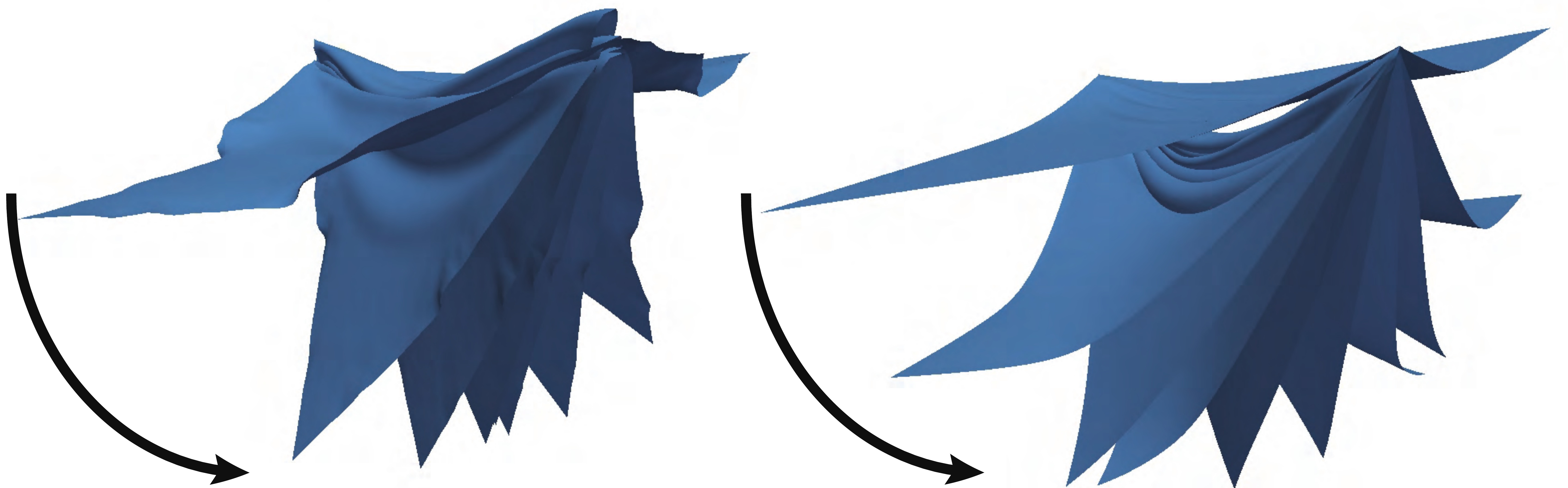}
\end{tabular}
\put(-100,93){(a)}
\put(-220,93){(b)}
\put(-100,-35){(c)}
\put(-220,-35){(d)}}
\end{center}
\vspace{-15px}
\caption{\label{fig:simulate} Two examples of simulation sequence generation in our feature space. (a): 5 frames in the simulation of a cloth swinging down. (b): Synthesized simulation sequence. (c): Another example where two diagonal points are grasped. (d): Synthesized simulation sequence.}
\vspace{-15px}
\end{figure}

For our second application, we approximate an entire cloth simulation sequence (\prettyref{eq:SIM}) in the 128-dimensional feature space. Starting from $\E{z}_1,\E{z}_2$, we can generate an entire sequence of $N$ frames by using the recurrent relationship in \prettyref{eq:MLP} and can recover the meshes via the function $\E{ACAP}^{-1}\circ\E{D}$. Such a latent space physics model has been previously proposed in \cite{1802.10123} for voxelized grids, while our model works on surface meshes. We show two synthesized simulation sequences in \prettyref{fig:simulate}. 

\subsection{Accuracy of Learned Simulator for Robotic Cloth Manipulation}
\begin{figure*}[ht]
\vspace{-10px}
\begin{center}
\scalebox{0.9}
{
\includegraphics[width=0.21\textwidth]{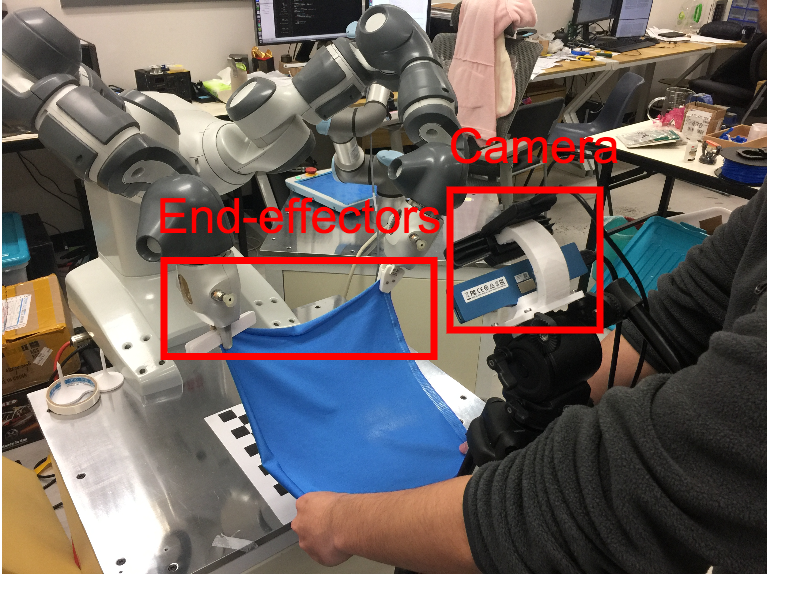}
\includegraphics[width=0.74\textwidth]{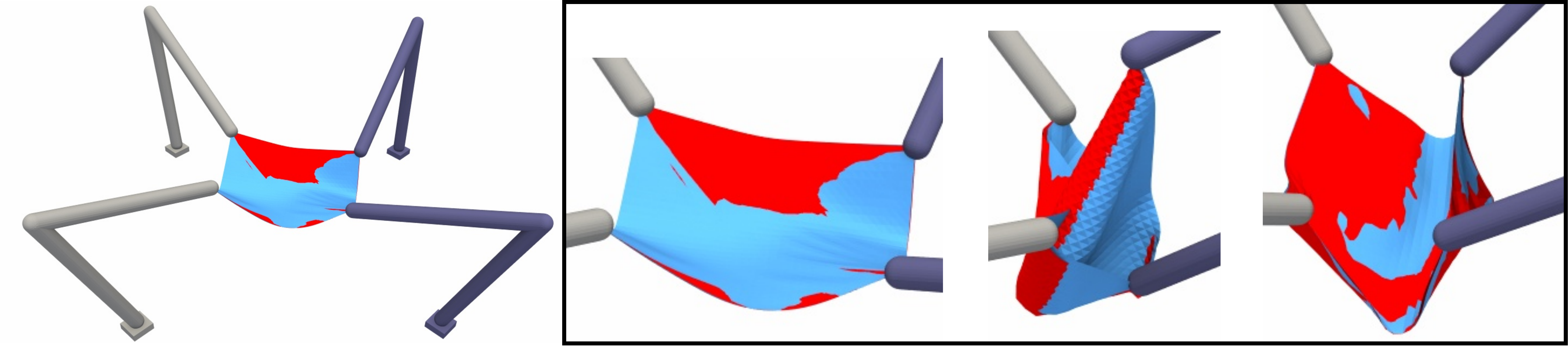}
\put(-370,70){(a)}
\put(-200,70){(b)}
\put(-120,70){(c)}
\put(-45 ,70){(d)}
}
\end{center}
\vspace{-10px}
\caption{\label{fig:simulateRobot} We reproduce benchmarks from \cite{1809.08259} where the robot is collaborating with human to manipulate a piece of cloth (a). We randomly perturb two grasp points on the left (gray arms) and the robot is controlling the other two grasp points (purple arms) using a visual-serving method to maintain the cloth at a target state, e.g., keeping the cloth flat (b), twisted (c), or bent (d). The red cloth is the groundtruth acquired by running the accurate FEM-based cloth simulator \cite{Narain:2012:AAR}, which takes 3 hours. The difference between our result (blue) and the groundtruth is indistinguishable.}
\vspace{-10px}
\end{figure*}

We show three benchmarks (\prettyref{fig:simulateRobot}) from robot cloth manipulation tasks defined in prior work \cite{1809.08259}. In these benchmarks, the robot is collaborating with human to maintain a target shape of a piece of cloth. To design such a collaborating robot controller, we use imitation learning by teaching the robot to recognized cloth shapes under various, uncertain human movements. Our learnable simulator can be used to efficiently generate these cloth shapes for training the controller. To this end, we train our neural-network using the original dataset from \cite{1809.08259} obtained by running the FEM-based simulator \cite{Narain:2012:AAR}, which takes 3 hours. During test time, we perturb the human hands' grasp points along randomly directions. Our learned physical model can faithfully predict the dynamic movements of the cloth.
\section{Results}\label{sec:results}
To evaluate our method, we create two datasets of cloth simulations using \prettyref{eq:PLOSS}. Our first dataset is called {SHEET}, which contains animations of a square-shaped cloth sheet swinging down under different conditions, as shown in \prettyref{fig:datasetVis} (a). This dataset involves $6$ simulation sequences, each with $N=2400$ frames. Among these $6$ sequences, the first sequence uses the mass-spring model \cite{Choi:2002:SBR:566570.566624} to discretize \prettyref{eq:PDE} and the cloth mesh has no holes (denoted as SHEET+\cite{Choi:2002:SBR:566570.566624}). \REFINED{The second and the third simulation sequences in the dataset use different material parameter, by multiplying the stretch/bend resistance term by $0.1$ and thereby making the material softer and less resilient when stretched or bent. These two sequences are denoted as SHEET+\cite{Choi:2002:SBR:566570.566624}+$0.1\mathcal{P}_s$ and SHEET+\cite{Choi:2002:SBR:566570.566624}+$0.1\mathcal{P}_b$, respectively.}
The forth sequence uses the mass spring model and the cloth mesh has holes, as shown in \prettyref{fig:simulate} (a,b), which is denoted as (SHEET+\cite{Choi:2002:SBR:566570.566624}+holes). The fifth sequence uses FEM \cite{Narain:2012:AAR} to discretize \prettyref{eq:PDE} and the cloth mesh has no holes (denoted as SHEET+\cite{Narain:2012:AAR}). The sixth sequence uses FEM to discretize \prettyref{eq:PDE} and the cloth interacts with an obstacle, as shown in \prettyref{fig:datasetVis} (c) (denoted as SHEET+\cite{Narain:2012:AAR}+obstacle). In the SHEET dataset, the cloth mesh without holes has $K=4225$ vertices and the cloth mesh with holes has $K=4165$ vertices. Our second dataset is called {BALL}, which contains animations of a cloth ball being dragged up and down under different conditions, as shown in \prettyref{fig:datasetVis} (d). This dataset also involves $4$ simulation sequences, each with $N=500$ frames. Using the same notation as the SHEET dataset, the $4$ sequences in the BALL dataset are (BALL+\cite{Choi:2002:SBR:566570.566624}, BALL+\cite{Narain:2012:AAR}, BALL+\cite{Narain:2012:AAR}+$0.1\mathcal{P}_s$, BALL+\cite{Narain:2012:AAR}+$0.1\mathcal{P}_b$). 
In the BALL dataset, the cloth ball mesh has $K=1538$ vertices. \REFINED{During comparison, for each dataset, we select first 12 frames in every 17 frames to form the training set. The other frames are used as the test set.}

\subsection{Implementation}
We implement our method using Tensorflow \cite{45381} and we implement the PB-loss as a special network layer. When there is an obstacle interacting with the cloth, we model the collision between the cloth and the obstacle using a special potential term proposed in \cite{gast2015optimization}. For better conditioning and a more robust initial guess, our training procedure is broken into three stages. During the first stage, we use the loss:
\begin{align*}
\mathcal{L}_1=\sum\lambda_i\mathcal{L}_i\quad {i\in\{recon,vert,ephys\}}
\end{align*}
to optimize $\E{E},\E{D}$. During the second stage, we use the loss:
\begin{align*}
\mathcal{L}_2=\sum\lambda_i\mathcal{L}_i\quad {i\in\{sim,mphys\}}
\end{align*}
to optimize $\E{MLP}$. Finally, we add a fine-tuning step and use the loss: 
\begin{align*}
\mathcal{L}_3=\sum\lambda'_i\mathcal{L}_i\quad {i\in\{recon,vert,ephys,sim\}}
\end{align*}
to optimize both $\E{E},\E{D}$ and $\E{MLP}$. Notice that, in order to train the mesh embedding network and $\E{MLP}$ at the same time, we feed:
\begin{align*}
\E{z}' = 0.5*\E{z}_m + 0.5*\E{MLP}(\E{z}_{m-2}, \E{z}_{m-1}, \E{q}_m)
\end{align*}
to $\E{D}$ for better stability during the third stage.

\subsection{Physics Correctness of Low-Dimensional Embedding}
We first compare the quality of mesh deformation embeddings using two different methods. \REFINED{The quality of embedding is measured using three metrics. The first metric is the root mean square error, $\mathcal{M}_{rms}$ \cite{doi:10.1111/j.1467-8659.2009.01602.x}, which measures the averaged vertex-level error over all $N$ shapes and $K$ vertices. Our second metric is the STED metric, $\mathcal{M}_{STED}$ \cite{Vasa:2011:PCC:1920053.1920227}. This metric linearly combines several aspects of errors crucial to visual quality, including relative edge length changes and temporal smoothness.} However, since $\mathcal{M}_{STED}$ is only meaningful for consecutive frames, we compute $\mathcal{M}_{STED}$ for the $5$ consecutive frames in every $17$ frames, which is the test set. Finally, we introduce a third metric, physics correctness, which measures how well the physics rule is preserved. Inspired by \prettyref{eq:VIO_MEASURE}, physics correctness is measured by the norm of partial derivatives of $\mathcal{L}_{phys}$: $\mathcal{M}_{phys}=\|\FPPROW{\mathcal{L}_{phys}}{\E{p}_m}\|^2$. Note that the absolute value of $\mathcal{M}_{phys}$ can vary case by case. For example, $\mathcal{M}_{phys}$ using the FEM method can be orders of magnitude larger than that using the mass-spring system in our dataset. So that only the relative value of $\mathcal{M}_{phys}$ indicates improvement in physics correctness.

\begin{table*}[t]
\setlength{\tabcolsep}{3pt}
\begin{center}
\begin{tabular}{cc}
\scalebox{0.9}{
\begin{tabular}{|c|c|c|c|c|}
\toprule
Dataset & Method & $\mathcal{M}_{rms}$ & $\mathcal{M}_{STED}$ & $\mathcal{M}_{phys}$	\\
\midrule
\multirow{3}{*}{SHEET+\cite{Choi:2002:SBR:566570.566624}} 
 & ours                & $\E{9.724}$ & $\E{0.01556}$ & $\E{14581.46}$ \\
\cline{2-5}
 & \cite{Tan2018AAAI}+$\mathcal{L}_{vert}$  & $10.351$ & $0.01688$ & $15446.70$ \\
\cline{2-5}
& \REFINED{\cite{Tan2018AAAI}}  & \REFINED{$11.841$} & \REFINED{$0.01760$} & \REFINED{$16198.02$} \\
\cline{1-5}
\multirow{2}{*}{\REFINED{\TWORCellC{SHEET+\cite{Choi:2002:SBR:566570.566624}}{+0.1$\mathcal{P}_s$}}} 
 & ours                & $\E{10.477}$ & $\E{0.01882}$ & $\E{1996.01}$ \\
\cline{2-5}
 & \cite{Tan2018AAAI}+$\mathcal{L}_{vert}$  & $11.341$ & $0.01957$ & $2128.05$ \\
\cline{1-5}
\multirow{2}{*}{\REFINED{\TWORCellC{SHEET+\cite{Choi:2002:SBR:566570.566624}}{+0.1$\mathcal{P}_b$}}} 
 & ours                & $\E{9.580}$ & $\E{0.01675}$ & $\E{10189.24}$ \\
\cline{2-5}
 & \cite{Tan2018AAAI}+$\mathcal{L}_{vert}$  & $11.319$ & $0.01773$ & $10260.39$ \\
\cline{1-5}
\multirow{2}{*}{\TWORCellC{SHEET+\cite{Choi:2002:SBR:566570.566624}}{+holes}} 
 & ours                & $\E{12.456}$ & $\E{0.02443}$ & $\E{19548.58}$ \\
\cline{2-5}
 & \cite{Tan2018AAAI}+$\mathcal{L}_{vert}$  & $12.928$ & $0.02480$ & $20632.13$ \\
\cline{1-5}
\multirow{2}{*}{SHEET+\cite{Narain:2012:AAR}} 
 & ours                & $\E{10.671}$ &  $0.01398$ & $\E{29109.03}$ \\
\cline{2-5}
 & \cite{Tan2018AAAI}+$\mathcal{L}_{vert}$  & $10.675$ & $\E{0.01195}$ & $103878.20$ \\
\bottomrule
\end{tabular}}
\scalebox{0.9}{
\begin{tabular}{|l|c|c|c|c|}
\toprule
Dataset & Method & $\mathcal{M}_{rms}$ & $\mathcal{M}_{STED}$ & $\mathcal{M}_{phys}$	\\
\midrule
\multirow{2}{*}{\TWORCellC{SHEET+\cite{Narain:2012:AAR}}{+obstacle}} 
 & ours                & $\E{8.075}$ & $\E{0.01784}$ & $160545.91$ \\
\cline{2-5}
 & \cite{Tan2018AAAI}+$\mathcal{L}_{vert}$  & $8.425$ & $0.01885$ & $\textcolor{blue}{\E{155117.28}}$ \\
\cline{1-5}
\multirow{3}{*}{BALL+\cite{Choi:2002:SBR:566570.566624}} 
 & ours                & $\E{17.008}$ & $\E{0.02570}$ & $\E{338465.20}$ \\
\cline{2-5}
& \cite{Tan2018AAAI}+$\mathcal{L}_{vert}$  & $20.326$ & $0.03018$ & $459564.89$ \\
\cline{2-5}
& \REFINED{\cite{Tan2018AAAI}}  & \REFINED{$22.798$} & \REFINED{$0.03622$} & \REFINED{$476210.46$} \\
\cline{1-5}
\multirow{2}{*}{BALL+\cite{Narain:2012:AAR}} 
 & ours                & $\E{11.663}$ & $\E{0.01501}$ & $\E{49392520.44}$ \\
\cline{2-5}
 & \cite{Tan2018AAAI}+$\mathcal{L}_{vert}$  & $12.200$ & $0.01662$ & $61048220.56$ \\
\cline{1-5}
\multirow{2}{*}{\REFINED{\TWORCellC{BALL+\cite{Narain:2012:AAR}}{+$0.1\mathcal{P}_s$}}} 
 & ours                & $\E{16.853}$ & $0.03394$ & $\E{80389.68}$ \\
\cline{2-5}
 & \cite{Tan2018AAAI}+$\mathcal{L}_{vert}$  & $17.706$ & $\E{0.03357}$ & $84508.12$ \\
\cline{1-5}
\multirow{2}{*}{\REFINED{\TWORCellC{BALL+\cite{Narain:2012:AAR}}{+$0.1\mathcal{P}_b$}}}
 & ours                & $\E{23.766}$ & $\E{0.04320}$ & $\E{674115.57}$ \\
\cline{2-5}
 & \cite{Tan2018AAAI}+$\mathcal{L}_{vert}$  & $27.390$ & $0.05320$ & $974481.48$ \\
\bottomrule
\end{tabular}}
\end{tabular}
\end{center}
\vspace{-10px}
\caption{\label{table:comparisonA} We compare the embedding quality using our method ($\lambda_{recon}=1, \lambda_{vert}=1, \lambda_{ephys}=0.1\ to\ 1$, where $\lambda_{ephys}$ for our method is tuned for different datasets.) and \cite{Tan2018AAAI}+$\mathcal{L}_{vert}$ ($\lambda_{recon}=\lambda_{vert}=1, \lambda_{ephys}=0$). \REFINED{We also compare results trained using dataset with different cloth material properties ($0.1\mathcal{P}_s$ means  that the stretch resilience has $1/10$ of the original material value   and $0.1\mathcal{P}_b$ means ten times less bending resilience).} From left to right: name of dataset, method used, $\mathcal{M}_{rms}$, $\mathcal{M}_{STED}$, and $\mathcal{M}_{phys}$.}
\vspace{-10px}
\end{table*}
Our first experiment compares the accuracy of mesh embedding with or without PB-loss. The version without PB-loss is our baseline, which is equivalent to adding vertex level loss to \cite{Tan2018AAAI}. In addition, we remove the sparsity regularization from \cite{Tan2018AAAI} to make it consistent with our formulation. We denote this baseline as \cite{Tan2018AAAI}+$\mathcal{L}_{vert}$. A complete summary of our experimental results is given in \prettyref{table:comparisonA}. The benefit of three-stage training is given in \prettyref{table:MLP}. From \prettyref{table:comparisonA}, we can see that including PB-loss significantly and consistently improves $\mathcal{M}_{phys}$. This improvement is large, up to $70\%$ on the {SHEET+\cite{Narain:2012:AAR}} dataset. In addition, by adding $\mathcal{L}_{ephys}$, our method also better recognizes the relationship between each model and embeds them, thus improves $\mathcal{M}_{rms}$ in all the cases. However, our method sometimes sacrifice $\mathcal{M}_{STED}$ as temporal smoothness is not modeled explicitly in our method. \REFINED{Finally, we have added two rows to \prettyref{table:comparisonA}, comparing our method with and without $\mathcal{L}_{vert}$, which shows that $\mathcal{L}_{vert}$ effectively reduces the error in terms of absolute vertex positions.}

\begin{figure}[ht]
\vspace{0px}
\begin{center}
\scalebox{0.8}{\includegraphics[width=0.47\textwidth]{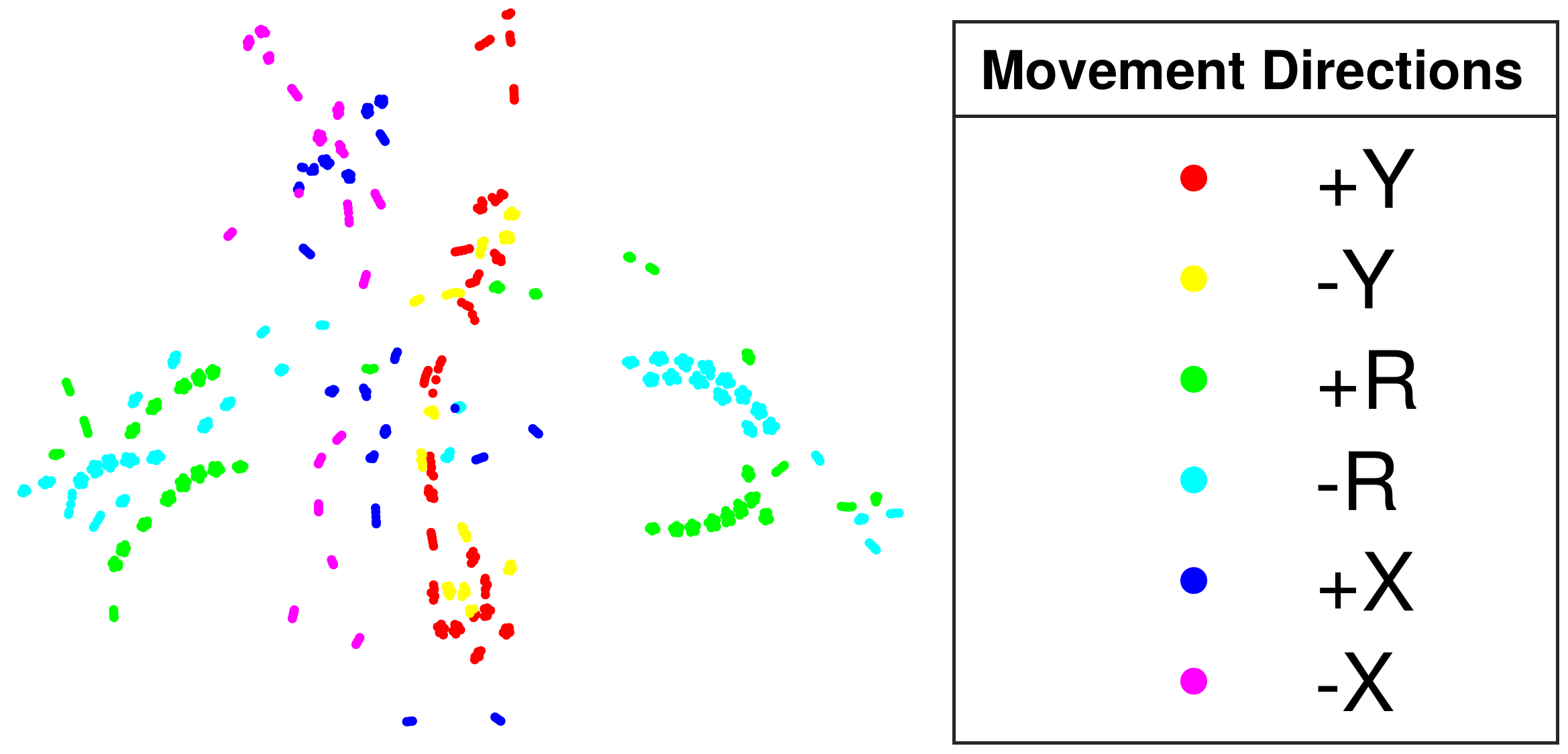}}
\end{center}
\vspace{-10px}
\caption{\label{fig:classify} A feature space visualization for SHEET+\cite{Choi:2002:SBR:566570.566624} using t-SNE.}
\vspace{-5px}
\end{figure}

\subsection{Discriminability of Feature Space}
In our second experiment, we evaluate the discriminability of mesh embedding by classifying the meshes using their feature space coordinates. Note that our datasets (\prettyref{fig:datasetVis}) are generated by moving the grasping points back and forth. We use these movement directions as the labels for classification. For the SHEET dataset, we have 6 labels: $\pm X/Y,\pm R$, where $\pm R$ means rotating the grasping points around $\pm Z$ axes. For the BALL dataset, we have 2 labels: $\pm Z$. Note that it is trivial to classify the meshes if we know the velocity of the grasping points. However, this information is missing in our feature space coordinates because ACAP features are invariant to global rigid translation, which makes the classification challenging. 
We visualize the feature space using t-SNE \cite{maaten2008visualizing} compressed to 2 dimensions in \prettyref{fig:classify}.
We report retrieval performance in the KNN neighborhoods across different K’s, using method suggested by \cite{yang2014optimization}. The normalized discounted cumulative gain (DCG) on the test set for SHEET+\cite{Choi:2002:SBR:566570.566624} is $0.8045$ and for BALL+\cite{Choi:2002:SBR:566570.566624} is $0.9128$.

\subsection{Sensitivity to Training Parameters}
In our third experiment, we evaluate the sensitivity of our method with respect to the weights of loss terms, as summarized in \prettyref{table:comparisonB}. Our method outperforms \cite{Tan2018AAAI}+$\mathcal{L}_{vert}$ under a range of different parameters. We have also compared our method with other baselines such as \cite{Neumann:2013:SLD:2508363.2508417} and \cite{huang}. As shown in the last two columns of \prettyref{table:comparisonB}, they generate even worse result, which indicates that \cite{Tan2018AAAI}+$\mathcal{L}_{vert}$ is the best baseline.
\begin{table}
\vspace{-5px}
\centering
\setlength{\tabcolsep}{5pt}
\scalebox{0.7}{
\begin{tabular}{|l|c|c|c|c|c|c|}
\toprule
Method & ($1$, $1$, $0.1$) & ($1$, $1$, $0.5$) & ($1$, $1$, $1$) & \cite{Tan2018AAAI}+$\mathcal{L}_{vert}$ & \cite{Neumann:2013:SLD:2508363.2508417} & \cite{huang}	\\
\midrule
$\mathcal{M}_{STED}$       & $0.01712$ & $\E{0.01635}$ & $\E{0.01556}$ & $0.01688$ & $0.03728$ & $0.02381$  \\
$\mathcal{M}_{phys}$       & $\E{15164.34}$ & $\E{14230.44}$ & $\E{14581.46}$ & $15446.70$ & $36038.51$ & $30732.07$ \\
\bottomrule
\end{tabular}}
\vspace{-2px}
\caption{\label{table:comparisonB} We compare the performance of our method with several previous ones in terms of $\mathcal{M}_{STED}$ and $\mathcal{M}_{phys}$ under different weights ($\lambda_{recon}$, $\lambda_{vert}$,  $\lambda_{ephys}$) of $\mathcal{L}$. The experiment is done on the dataset SHEET+\cite{Choi:2002:SBR:566570.566624}. Our method outperforms \cite{Tan2018AAAI}+$\mathcal{L}_{vert}$ over a wide range of parameters. Previous methods, including \cite{Neumann:2013:SLD:2508363.2508417} and \cite{huang}, generate even worse results, which supports our choice of using convolutional neural network and ACAP feature for mesh deformation embedding.}
\vspace{-5px}
\end{table}

\subsection{Robustness to Mesh Resolutions}
In this experiment, we highlight the robustness of our method to different mesh resolutions by lowering the resolution of our dataset. For SHEET+\cite{Choi:2002:SBR:566570.566624}, we create a mid-resolution counterpart with $K=1089$ vertices and a low-resolution counterpart with $K=289$ vertices. On these two new datasets, we compare the accuracy of mesh embedding with or without PB-loss. The results are given in \prettyref{table:comparisonC}. Including PB-loss consistently improves $\mathcal{M}_{phys}$ and overall embedding quality, no matter the resolution used.
\begin{table}[t]
\vspace{-5px}
\centering
\scalebox{0.9}{
\setlength{\tabcolsep}{3pt}
\begin{tabular}{|l|c|c|c|c|c|}
\toprule
Dataset & \#Vertices & Method & $\mathcal{M}_{rms}$ & $\mathcal{M}_{STED}$ & $\mathcal{M}_{phys}$	\\
\midrule
\multirow{6}{*}{SHEET+\cite{Choi:2002:SBR:566570.566624}} 
 & 4225  & ours                                    & $\E{9.724}$ & $\E{0.01556}$ & $\E{14581.46}$ \\
\cline{3-6}
 & 4225  & \cite{Tan2018AAAI}+$\mathcal{L}_{vert}$ & $10.351$ & $0.01688$ & $15446.70$ \\
\cline{3-6}
 & 1089  & ours                                    & $\E{11.744}$ & $\E{0.01511}$ & $\E{15712.89}$ \\
\cline{3-6}
 & 1089  & \cite{Tan2018AAAI}+$\mathcal{L}_{vert}$ & $11.842$ & $0.1648$ & $16076.44$ \\
\cline{3-6}
 & 289   & ours                                    & $\E{11.757}$ & $\E{0.01273}$ & $\E{10565.27}$ \\
\cline{3-6}
 & 289   & \cite{Tan2018AAAI}+$\mathcal{L}_{vert}$ & $12.510$ & $0.01543$ & $15498.58$ \\
\bottomrule
\end{tabular}}
\vspace{-5px}
\caption{\label{table:comparisonC} We profile the improvement in various metrics under different mesh resolution ($\lambda_{recon}=1, \lambda_{vert}=1, \lambda_{ephys}=0.5\ to\ 1$), compared with \cite{Tan2018AAAI}+$\mathcal{L}_{vert}$. From left to right: name of dataset, number of vertices, method used, $\mathcal{M}_{rms}$, $\mathcal{M}_{STED}$, and $\mathcal{M}_{phys}$. Our method consistently outperforms \cite{Tan2018AAAI}+$\mathcal{L}_{vert}$.}
\vspace{-5px}
\end{table}

\begin{table}[t]
\vspace{-0px}
\centering
\scalebox{0.9}{
\setlength{\tabcolsep}{2pt}
\begin{tabular}{|l|c|c|c|c|}
\toprule
Dataset & Method & $\mathcal{M}_{rms}$ & $\mathcal{M}_{STED}$ & $\mathcal{M}_{phys}$	\\
\midrule
\multirow{3}{*}{SHEET+\cite{Choi:2002:SBR:566570.566624}} 
 & baseline & $15.184$ & $0.01765$ & $21019.28$ \\
\cline{2-5}
 & 2nd stage & $14.910$ & $0.01705$ & $18103.66$ \\
 \cline{2-5}
 & 3rd stage & \small{$\E{14.000}$} & \small{$\E{0.01672}$} &\small{$\E{17819.85}$}\\
\midrule
\multirow{3}{*}{SHEET+\cite{Choi:2002:SBR:566570.566624}+holes} 
 & baseline & $18.843$ & $0.02703$ & $29673.30$ \\
\cline{2-5}
 & 2nd stage & $17.909$ & $0.02667$ & $28526.89$ \\
 \cline{2-5}
 & 3rd stage & \small{$\E{17.412}$} & \small{$\E{0.02633}$} & \small{$\E{28393.81}$} \\
\bottomrule
\end{tabular}}
\vspace{-5px}
\caption{\label{table:MLP} We compare the physical simulation performance of $\E{MLP}$ after training with $\lambda_{mphys} = 0$ (baseline), training with $\lambda_{mphys} = 0.5\ to\ 1$ (2nd stage), and fine-tuning (3rd stage). For the $5$ consecutive meshes in every $17$ frames (the test set), we give $\E{MLP}$ the first $2$ frames and predict the remaining $3$ frames to generate this table.}
\vspace{-5px}
\end{table}

\begin{table}[h]
\vspace{-5px}
\centering
\setlength{\tabcolsep}{2pt}
\scalebox{0.9}{
\begin{tabular}{|c|c|c|c|c|}
\toprule
Dataset & Method & $\mathcal{M}_{rms}$ & $\mathcal{M}_{STED}$ & $\mathcal{M}_{phys}$    \\
\midrule
\multirow{2}{*}{SHEET+\cite{Choi:2002:SBR:566570.566624}} 
 & FC+$\mathcal{L}_{vert}$+$\mathcal{L}_{ephys}$                  & $\E{22.700}$ & $\E{0.01742}$ & $\E{16565.55}$ \\
\cline{2-5}
 & FC+$\mathcal{L}_{vert}$  & $24.020$ & $0.02043$ & $19628.23$ \\
\cline{1-5}
\multirow{2}{*}{\TWORCellC{SHEET+\cite{Choi:2002:SBR:566570.566624}}{+holes}} 
 & FC+$\mathcal{L}_{vert}$+$\mathcal{L}_{ephys}$               & $\E{18.513}$ & $\E{0.02504}$ & $\E{21852.67}$ \\
\cline{2-5}
 & FC+$\mathcal{L}_{vert}$     & $19.253$ & $0.02684$ & $22748.91$ \\
\bottomrule
\end{tabular}
}
\caption{\REFINED{We use a fully connected underlying neural network like \cite{Fulton:LSD:2018} and train with or without our PB-loss. The profiled results show that our method can improve performance in terms of $\mathcal{M}_{rms}$, $\mathcal{M}_{STED}$, $\mathcal{M}_{phys}$, which is independent of the type of neural network architectures.}}
\label{table:fc}
\vspace{-5px}
\end{table}

\REFINED{\subsection{PB-Loss with Alternative Neural Network Architectures}
Our PB-loss is orthogonal to the architecture of neural networks. Therefore, we have conducted an additional set of experiments to highlight the performance improvement using a full-connected underlying neural network like \cite{Fulton:LSD:2018}. The results are shown in \prettyref{table:fc}.}

\subsection{Difficulty in Contact Handling}
One exception appears in the {SHEET+\cite{Narain:2012:AAR}+obstacle} (blue row in \prettyref{table:comparisonA}), where our method deteriorates physics correctness. This is the only dataset where the mesh is interacting with an obstacle. The deterioration is due to the additional loss term penalizing the penetration between the mesh and the obstacle. This term is non-smooth and has very high value and gradient when the mesh is in penetration, making the training procedure unstable. This means that direct learning a feature mapping for meshes with contacts and collisions can become unstable. However, we can solve this problem using a two-stage method, where we first learn a feature mapping for meshes without contacts and collisions, and then handle contacts and collisions at runtime using conventional method \cite{hirota2000simulation}, as is done in \cite{Barbic:2010:SSC}.

\subsection{Speedup over FEM-Based Cloth Simulators}
\REFINED{In \prettyref{table:RUNNING_TIME}, we have used our  neural network as a physics simulator and compared its performance with a conventional FEM-based method. Our method achieves $500-10000\times$ speedup over the FEM-based method \cite{Narain:2012:AAR} on average.}
\begin{table}[t]
\centering
\scalebox{0.9}{\setlength{\tabcolsep}{3pt}
\begin{tabular}{|l|c|c|c|}
\toprule
Dataset & \#Vertices & NN(s) & FEM(s)	\\
\midrule
SHEET+\cite{Choi:2002:SBR:566570.566624} & 4225 & 3.259  & 9094.906  \\
SHEET+\cite{Choi:2002:SBR:566570.566624}+holes & 4166 & 3.432  & 10336.001  \\
\prettyref{fig:simulateRobot} (b) & 1089 & 1.697 & 582.635  \\
\prettyref{fig:simulateRobot} (c) & 1089 & 1.701 & 701.971  \\
\prettyref{fig:simulateRobot} (d) & 1089 & 1.883 & 893.197  \\
\bottomrule
\end{tabular}}
\vspace{-5px}
\caption{\label{table:RUNNING_TIME} We compare the computational time to generate a sequence of cloth data using our stateful neural network and the conventional FEM-based method.}
\vspace{-5px}
\end{table}
\section{Conclusion \& Limitations}\label{sec:conc}
In this paper, we present a new method that bridges the gap between mesh embedding and and physical simulation for efficient dynamic models of clothes. We achieve low-dimensional mesh embedding using a stateless, graph-based CNN that can handle arbitrary mesh topologies. To make the method aware of physics rules, we augment the embedding network with a stateful feature space simulator represented as a MLP. The learnable simulator is trained to minimize a physics-inspired loss term (PB-loss). This loss term is formulated on the vertex level and the transformation from the ACAP feature level to the vertex level is achieved using the inverse of the ACAP feature extractor.

Our method can be used for several applications, including fast inverse kinematics of clothes and realtime feature space physics simulation. We have evaluated the accuracy and robustness of our method on two datasets of physics simulations with different material properties, mesh topologies, and collision configurations. Compared with previous models for embedding, our method achieves consistently better accuracy in terms of physics correctness and the mesh change smoothness metric (\cite{Vasa:2011:PCC:1920053.1920227}).

A future research direction is to apply our method to other kinds of deformable objects, i.e., volumetric objects \cite{volumeManip}. Each and every step of our method can be trivially extended to handle volumetric objects by replacing the triangle surface mesh with a tetrahedral volume mesh. A minor limitation of the current method is that the stateful MLP and the stateless mesh embedding cannot be trained in a fully end-to-end fashion. We would like to explore new optimization methods to train the two networks in an end-to-end fashion while achieving good convergence behavior. \REFINED{Finally, our approach is limited to a single setting of thin-shell simulation and needs to be re-trained whenever there are changes in the resolution of the mesh, the material parameters, or the obstacles in the environment.}

\section{Acknowledgement}
This work was supported by ARO grants (W911NF1810313 and  W911NF1910315) and Intel. Lin Gao was partially supported by the National Natural Science Foundation of China (No. 61872440) and Beijing Municipal Natural Science Foundation (No. L182016).
\bibliographystyle{IEEEtranS}
\bibliography{deform}
\end{document}